\documentclass[twocolumn]{article}

\usepackage[big]{dgruyter}
\usepackage{rotating}

\usepackage{longtable}
\usepackage{pdflscape}
\usepackage{translations}
\usepackage{romannum}
\usepackage{graphicx}	
\usepackage{amsmath}	
\usepackage{amssymb}	
\usepackage{multirow}
\usepackage{bm}
\usepackage{authblk}

\usepackage{color}
\usepackage{caption}
\setcitestyle{authoryear,open={(},close={)}}

\newcommand\kms{\ensuremath{\mbox{km}\,\mbox{s}^{-1}}}
\newcommand\Teff{\ensuremath{T_\mathrm{eff}}}
\newcommand\logg{\ensuremath{\log g}}
\newcommand\vsini{\ensuremath{v_{e}\sin i}}

\newcommand\met{\ensuremath{[M/H]}}

\begin{document}

\author*[1]{Marwan Gebran$^1$}
\author[3]{Fr\'ed\'eric Paletou$^2$}
\author[4]{Ian Bentley$^1$}
\author[2]{Rose Brienza$^1$}
\author[2]{Kathleen Connick$^1$}

\runningauthor{M. Gebran}  
\affil[1]{Department of Chemistry and Physics, Saint Mary’s College, Notre Dame, IN 46556, USA, E-mail: mgebrane@saintmarys.edu}

\affil[2]{Universit\'e de Toulouse, Observatoire Midi--Pyr\'en\'es, Irap, Cnrs, Cnes,  14 av. E. Belin, F--31400 Toulouse, France}

\baretabulars 

\articletype{Research Article}

\title{Deep Learning application for stellar parameters determination: \\
II- Application to observed spectra of AFGK stars}
\runningtitle{DL for stellar parameters II}

\begin{abstract}
{In this follow-up paper, we investigate the use of Convolutional Neural Network for deriving stellar parameters from observed spectra. Using hyperparameters determined previously, we have constructed a Neural Network architecture suitable for the derivation of \Teff, \logg, \met, and \vsini. The network was constrained by applying it to databases of AFGK synthetic spectra at different resolutions. Then, parameters of A stars from Polarbase, SOPHIE, and ELODIE databases are derived as well as FGK stars from the Spectroscopic Survey of Stars in the Solar Neighbourhood. The network model average accuracy on the stellar parameters are found to be as low as 80 K for \Teff, 0.06 dex for \logg, 0.08 dex for \met, and 3 km/s for \vsini\ for AFGK stars.} 
\end{abstract}

 \journalname{Open Astronomy}
\DOI{DOI}
  \startpage{1}
  \received{..}
  \revised{..}
  \accepted{..}

  \journalyear{2022}
  \journalvolume{1}

\maketitle

\section{Introduction}
Artificial Intelligence (AI) is becoming a vital tool in science due to its automation capabilities and its capacity to handle large amounts of data. In the context of astronomy, a subset of AI, Machine and Deep Learning (ML and DL) are used extensively for ground-based and sky surveys \citep{2019arXiv190407248B}. In our previous work, \cite{Gebran22} (hereafter referred to as Paper I), we constructed a Deep Neural Network (DNN) in order to derive stellar parameters\footnote{When dealing with DNN, stellar parameters are often called stellar labels.} such as: effective temperature (\Teff), surface gravity (\logg), metallicity (\met), and the projected equatorial rotational velocity (\vsini) for B and A stars. In Paper I, we constrained most of the hyperparameters utilized in the construction of the Neural Network (NN) in order to insure that the best accuracy for deriving stellar labels could be achieved.

Many tools and techniques are being developed to derive the fundamental parameters of stars and most of them are either based on statistical or ML/DL approaches. A thorough list of the most updated studies can be found in the introduction of Paper I. Recently, \cite{2022arXiv220406301L} used a combination of Least Absolute Shrinkage and Selection Operator (LASSO) and Multi-layer Perceptron (MLP) methods 
to estimate stellar atmospheric parameters from the Large Sky Area Multi-Object Fiber Spectroscopic
Telescope (LAMOST) DR8 low-resolution spectra. \cite{2022AJ....163..236S} presented a spectral analysis algorithm, ZETA-PAYNE, developed to obtain stellar labels from SDSS-V spectra of stars of OBAF spectral types using machine learning
tools. \cite{2022MNRAS.511.5597L} applied a ML technique, the Gaussian Process (GP) regression, to turn a sparse
model grid into a continuous function. They also used the GP regression to determine the age and mass of stars. \cite{2021arXiv211109081K} presented a neural network autoencoder approach for extracting a telluric transmission spectrum from a large set of high-precision observed solar spectra from the The High Accuracy Radial Velocity Planet Searcher (HARPS-N) radial velocity spectrograph. \cite{2022ApJ...930...70H} presented a data-driven method based on long short-term memory (LSTM) neural networks to analyze spectral time series of Type Ia supernovae (SNe Ia). Their method allows for accurate reconstruction of the spectral sequence of an SN Ia based on a single observed spectrum around maximum light. More recently, \cite{RRnet} presented a Residual Recurrent Neural Network (RRNet) to extract spectral information, and estimate stellar atmospheric parameters along with 15 chemical element
abundances for medium-resolution spectra from LAMOST.

Most of the automated techniques that are found in the literature deal with the derivation of the fundamental parameters (\Teff, \logg, \met) without considering \vsini. Rotational profiles are usually found by applying transformations such as the Fourier Transform (FT, \citealt{2012A&A...537A.120Z}) or using rotational laws \citep{2017A&A...602A..83Z}. In our previous studies \citep{2014sf2a.conf..451A,Gebran,2019OAst...28...68K,Gebran22}, we have derived \vsini\ using line profile fitting. 

In this work, we complement the study of Paper I by using its best combination of hyperparameters to find the best NN architecture. 
The foremost purpose of our study is to develop a consistent model that is capable of predicting accurate and precise stellar parameters which is a guiding starting point to most stellar physics projects. Of course other sources of uncertainties can affect the predicted results when applied to real observations as will be discussed in this paper.  Once the architecture and parameters are set, our technique is then applied to AFGK observed spectra \citep{Gebran,2019OAst...28...68K,S4n}. 

The construction of the training databases is explained in Sec.~\ref{training}. The pre-processing steps are detailed in Sec.~\ref{pca}. The construction of the NN model with all the details are discussed in Sec.~\ref{dl}. The application of the method to AFGK stars is found in Secs.~\ref{observation}. The discussion and conclusion can be found in Sec.~\ref{discussion}.

\section{Training Databases}
\label{training}
A grid of 12 training databases was constructed for the purpose of this study. Other than modifying the stellar labels (\Teff, \logg, \met\footnote{\met\ refers to an overall metallicity and not to the iron abundance. All elemental abundances are scaled according to \met.}, \vsini), we have constructed a series of grids with the same range of stellar labels but at different resolving power. The purpose is to analyse the effect of resolution on the accuracy of the derived stellar label and to check the capability of our technique when applied to low, medium, and high resolution spectrometers.

\begin{figure*}[!h]
    \centering
\includegraphics[scale=0.35]{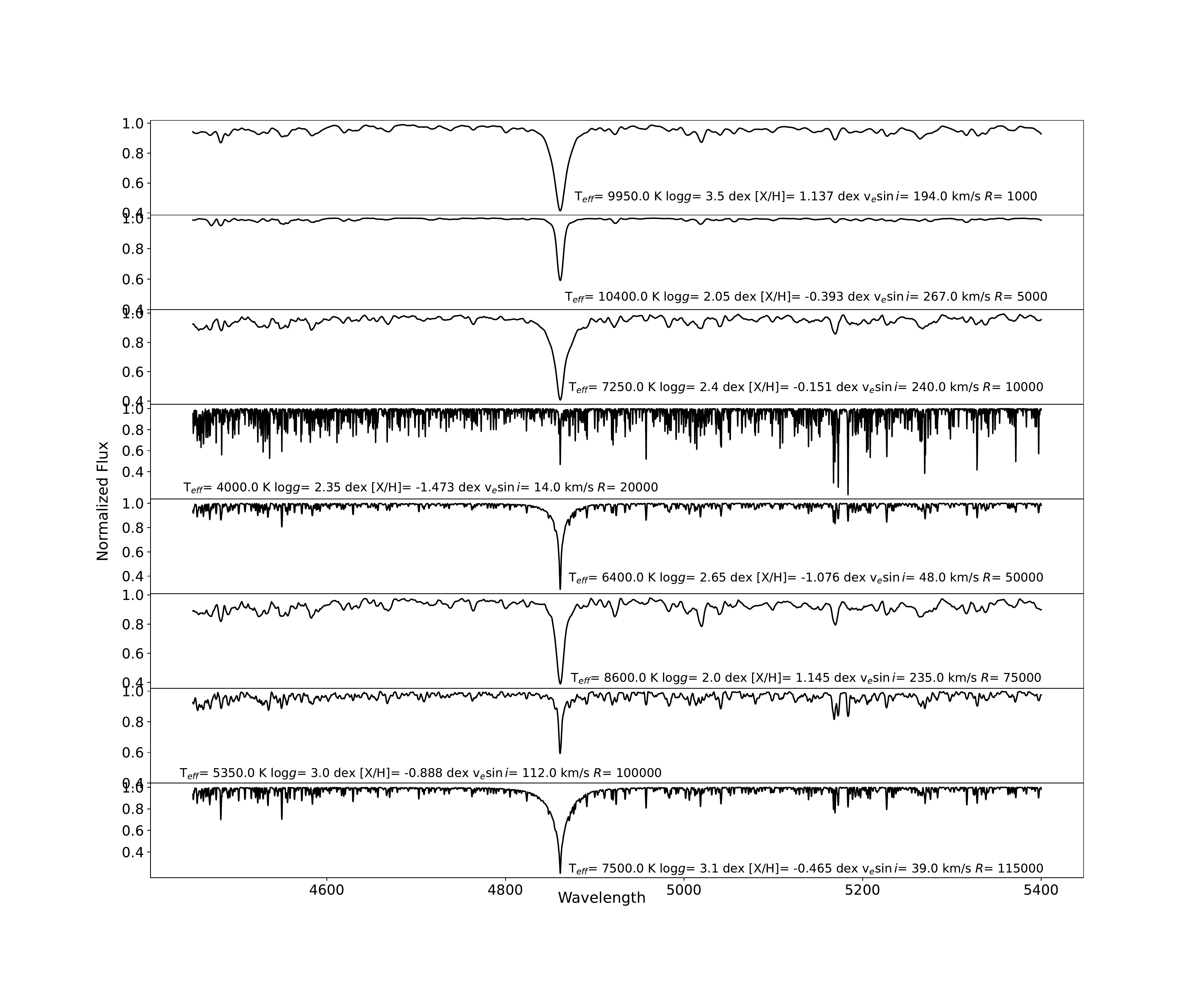}
    \caption{Sample of synthetic spectra calculated with different stellar parameters and spectral resolutions. These noise free spectra are normalized to the local continuum.} 
    \label{synthetic}
\end{figure*}

We have followed the same strategy as in Paper I. We have first calculated a series of \texttt{ATLAS9} \citep{Kurucz1992} model atmospheres using the opacity distribution function of \cite{castelli} and with a mixing length  parameter of 0.5 for 7000 K$\leq$\Teff$\leq$8500 K, and 1.25 for \Teff$\leq$7000 K \citep{2004IAUS..224..131S}. Using \texttt{SYNSPEC48} \citep{spectra}, we have calculated synthetic spectra for AFGK stars. We ended up, for each resolving power, with a grid of 80\,000 synthetic spectra with parameters ranging randomly between the values described in Tab.~\ref{rangeofparameters}. We have used the same line list as the one used in Paper I. The wavelength range was selected to be between 4450 and 5400 \AA. This wavelength range is indeed very sensitive to all stellar parameters in the spectral range of AFGK stars \citep{S4n, Gebran, 2019OAst...28...68K, Gebran22}. This region is also insensitive to microturbulent velocity which was adopted to be $\xi_t$=2 km/s for A stars and $\xi_t$=1 km/s for FGK stars \citep{Gebran,micro}.
A example of the calculated spectra for different resolving powers is displayed in Fig.~\ref{synthetic}.

\begin{table*}[!thb]
\centering
\caption{Ranges of the parameters used for the calculation of the training databases. The $3^{rd}$ column displays the steps in the parameter range. Note that the steps in \Teff\ and \logg\ are the steps in \texttt{ATLAS9} model atmospheres. Many databases were constructed for different resolving power ranging from 1\,000 to 115\,000. Random steps means that there is no restriction on the number selection.}
\label{rangeofparameters}
\begin{tabular}{|c|c|c|} 
 \hline
	Parameters & Range& Step \\
	\hline
	\Teff\ (K) & [4\,000,11\,000] & 50  \\
	\logg\ (dex)  & $[2.0, 5.0]$ & 0.05 \\
	\met   (dex)  & $[-1.5 , 1.5]$  & random \\ 
	\vsini\ (\kms) & $[0, 300] $  & random\\
	$\lambda$ (\AA)& 4\,450-5\,400 & $\dfrac{\lambda}{\mathrm{Resolving \ \ Power}}$ \\

 	\hline
  \end{tabular}
\end{table*}

\section{PCA for Pre-processing}
\label{pca}
Before applying the NN to the training database, a dimension reduction technique is applied. This step consists in reducing the size of the spectra from a sampling size of $N_{\lambda}$ to $n_k<N_{\lambda}$. Depending on the resolving power, $N_{\lambda}$ ranges between 4\,750 to 19\,000 data points. The $n_k$ reduced data points are found by projecting the flux of each spectrum onto the first 50 Principal Components (PC). Technically, we apply this Principal Component Analysis (PCA) on the training database and we find the eigenvectors $\textbf{e}_k(\lambda)$ of the variance-covariance matrix $\textbf{\textit{C}}$:

\begin{equation}
\textbf{\textit{C}}=(\textbf{\textit{M}}-\bar{\textit{M}})^\mathrm{T}\cdot(\textbf{\textit{M}}-\bar{\textit{M}}) \, , 
\end{equation}
where the training database $\textbf{\textit{M}}$ is a $N_{\mathrm{spectra}} \times N_{\lambda}$ matrix containing the fluxes of the synthetic spectra. The value of $n_k$ is found by analyzing the reconstructed error \citep{Gebran22}. Having $n_k$=50 reduces the mean reconstructed error to a value less than 0.5\%. As a similar technique to PCA, one could also use the encoder part of an autoencoder in order to reduce the dimensonality of the database \citep{2021arXiv211109081K}. We have chosen PCA to be consistent with our previous findings in \cite{Gebran} and \cite{Gebran22}.

\section{Deep Learning}
\label{dl}
 We start by applying data augmentation as a regularization technique to all the training databases (see Sec.~4.1.1 of Paper I for technical explanations). This is done in order to take into account the noise in the real observed spectra as well as considering some modifications that could occur in the shape of the observed spectra due to a bad normalization or inappropriate data reduction. Every spectrum (including the augmented ones) in each database is represented by 50 data points and they correspond to a specific \Teff, \logg, \met, and \vsini. This is true at all resolving powers. A NN is then used to link these data points to their corresponding labels. Four different NNs were built, one for each stellar label (\Teff, \logg, \met, \vsini). 

The initializers, optimizers, learning rates, dropout fraction, pooling layers, activation functions, loss functions, epochs, and batches are constrained according to the methodology of Paper I. These network parameters were derived for every network architecture tested in this work. 

\subsection{Architecture}
\label{architecture}
An infinite number of architectures could be applied to our purpose. The main goal is to find the most accurate transformation between the matrix of spectra coefficients (the 50 projected ones) and the labels. The best architecture will be selected according to its simplicity (size and calculation time) and to the accuracy of the results.

Fully dense NNs, Convolutional Neural Networks (CNN), and a combination of both were tested for each stellar label. In each case we have iterated on the number of layers, number of neurons in each layer, and the size of the filters in case of CNNs. As explained previously, network parameters were derived for each NN.

For every network and every resolving power, each augmented database was divided in 70\% for training, 20\% for validation, and 10\% for testing. Gaussian signal to noise ratio (SNR) was selected randomly between 5 and 300 and applied to each spectrum of the  10\% test spectra in order to check the accuracy of the technique on noisy data.

All our calculations are performed using the open-source programming language, \texttt{Python}, specifically with the \texttt{Keras}\footnote{\url{https://keras.io/}} interface on the TensorFlow \texttt{TensorFlow}\footnote{\url{https://www.tensorflow.org/}}. We have used the \texttt{KerasTuner}\footnote{\url{https://keras.io/keras_tuner/}} package \citep{omalley2019kerastuner}, a scalable hyperparameter optimization framework that solves the pain points of hyperparameter search. It was used to derive the optimized number of layers as well as the filter sizes in case of CNNs. Linking the number of layers and dimension of the filters with the size of the database as well as the size of each spectrum in the database is not an easy task. In order to avoid over- and under-fitting, these two parameters should be optimized. \texttt{KerasTuner} helps in that regard and avoids the hassle of the time consuming trial and error phase.

After iterating the architecture shape and deriving the optimized parameters for each network, the result was a unique architecture that is applicable to all stellar parameters. Figure \ref{architecture} shows the architecture of NN for deriving \Teff. This architecture is similar for predicting \logg, \met, and \vsini. Although the four parameters are predicted with networks having similar architecture, these models differ in the activation function, the kernel initializer, the loss function, the optimizer, the dropout fraction, the epochs and the batches number \citep{Gebran22}. The adopted values for the network parameters are derived using the technique explained in Paper I. These parameters are summarized in Tab.~\ref{hyperparameters}.

\begin{figure}[!htp]
    \centering
    \includegraphics[scale=0.27]{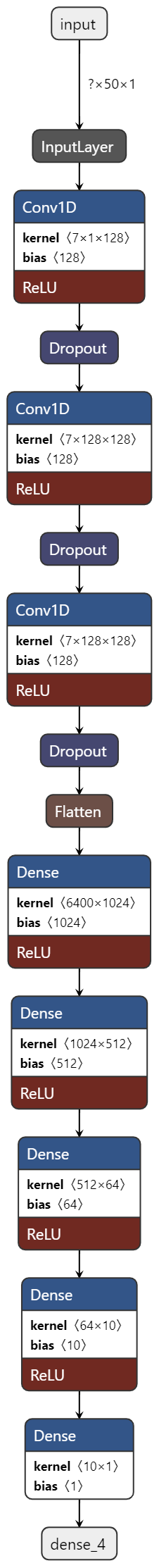}
    \caption{Neural Networks architecture used in this study. The parameters displayed in this model are the ones used for predicting \Teff,  \logg, \met, and \vsini\ are predicted with networks having the same architecture but different parameters as shown in Tab.~\ref{hyperparameters}. Explanations about the kernel and bias dimensions can be found in \cite{wu2017introduction}.}
    \label{architecture}
\end{figure}

\begin{table*}[]
\small
    \centering
        \caption{Set of parameters used for the 4 networks, for deriving \Teff, \logg, \met, and \vsini. These parameters are derived using the technique explained in Paper I.}

    \begin{tabular}{|l|c|c|c|c|}
    \hline
    Parameter& \Teff & \logg & \met & \vsini \\ \hline
    
Kernel initializer& he\_normal & he\_normal& Random Uniform & he\_uniform \\
Loss function & mean squared logarithmic error& mean squared logarithmic error & mean absolute error & mean squared error\\ 
Optimizer &Adam & Adamax & Adam & Adamax \\
Epochs& 350& 75 & 75 & 75\\
Batch&128& 128 &32 & 64 \\
Activation function& Relu & tanh & tanh & tanh \\
Dropout fraction&0.3& 0.3 & 0.2 & 0.3 \\
\hline
    \end{tabular}
    \label{hyperparameters}
\end{table*} 

\subsection{Resolution Effect}
\label{res}
Spectroscopic surveys are based on instruments that have different resolving power. For that reason, we have applied our technique to different databases that are similar in parameter ranges (Tab.~\ref{rangeofparameters}), except for the resolving power. Our tests contain spectra having a low resolution down to 1\,000 and a high resolution up to 115\,000. 

\begin{figure}[!b]
    \centering
\includegraphics[scale=0.42]{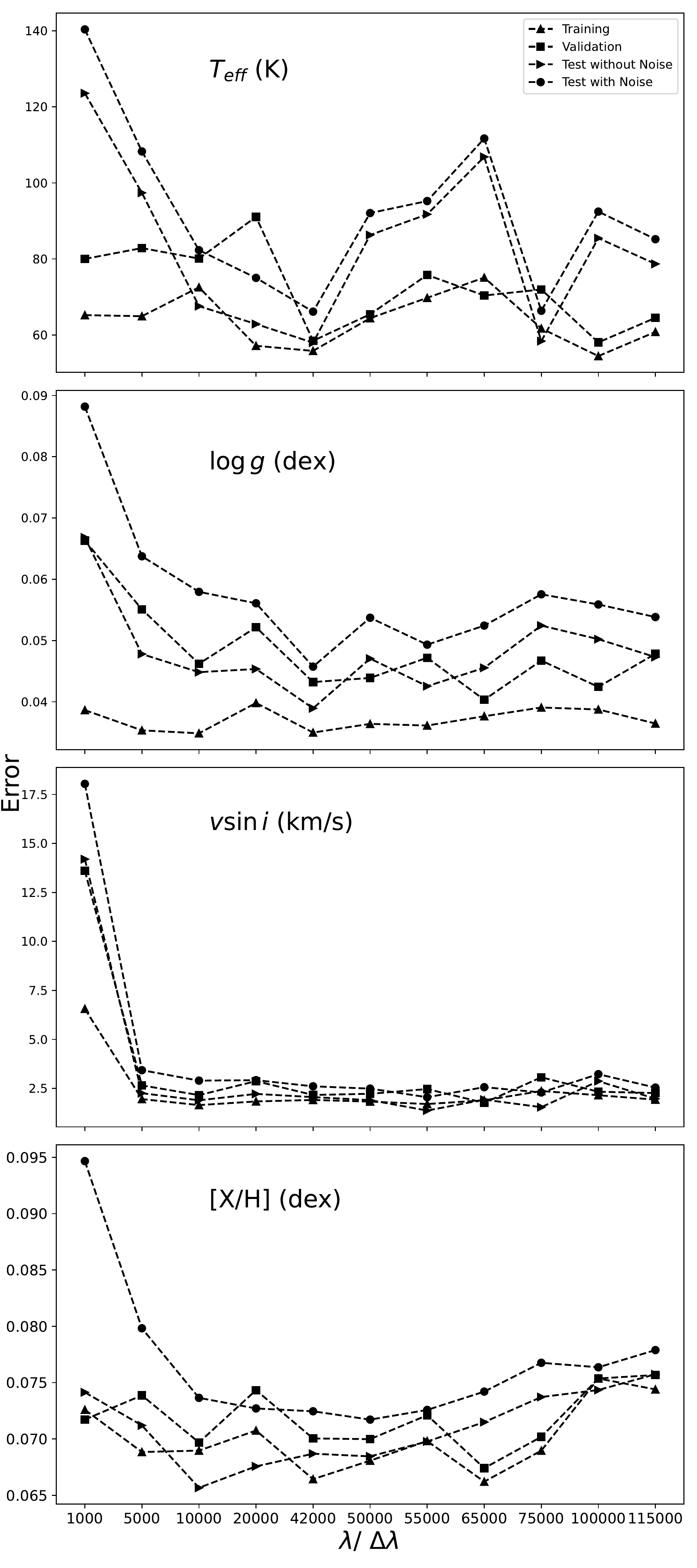}
    \caption{Derived accuracy for \Teff, \logg, \met, and \vsini\ as a function of the resolving power. We present the accuracy's for the training data (triangles), the validation (squares), test (triangles tilted right), and noisy test (circles) data.}
    \label{resolution}
\end{figure}

Once the networks are trained using the 70\% of the data, we have derived the accuracy on the parameters for the validation, test, and noisy test data. The best accuracy reached as a function of the resolution are displayed in Fig.~\ref{resolution}. For each stellar label, the derived accuracy for the noisy test data is representative of the error bar that should be assigned to observed spectra. For example, analyzing spectra at a resolving power of 50\,000, the equatorial projected rotational velocity should be assigned and error $\sigma_{\vsini_m}\sim$2.5 km/s. The subscript $m$ corresponds to the fact that this is model related. For a resolving power larger than 5\,000, the accuracy's are always in the same order and their average is  80 K for \Teff, 0.06 dex for \logg, 0.08 dex for \met, and 3 km/s for \vsini.

\section{Application to Observed Spectra}
\label{observation}
After the four networks were applied to synthetic data and the architecture and parameters found, we used them to predict the stellar parameters from \emph{observed} spectra. We used well studied AFGK stars observed with different instruments at different resolution. Applying the predictions to observed spectra assumes that the radiative transfer code is able to produce synthetic spectra similar to the observed ones using the specific stellar parameters. We have shown in previous studies \citep{Gebran,2019OAst...28...68K} that \texttt{SYNSPEC48} was able to reproduce spectra of AFGK stars with good accuracy but other reliable synthetic spectra codes could be used if needed. We can mention the \texttt{PHOENIX} models \citep{2013A&A...553A...6H} that are well suited for stars having \Teff\ $\leq$ 12\,000 K or TURBOSPECTRUM \citep{2012ascl.soft05004P} with all the molecular data that is used for giants and dwarfs stars. 

For the A stars, we used the list of \cite{Gebran} and selected the ones that have the most values published in the literature. We ended up with 89 observed A stars with more than 9 values for \Teff\ in the the Vizier catalog. These A stars were observed with NARVAL and ESPaDOnS high-resolution spectropolarimeters and having a spectral resolving power of 65\,000 in polarimetric mode and 76\,000 when used for spectroscopy. These spectra were retrieved from PolarBase \citep{polar}, a high resolution spectropolarimetric stellar observations database. Another part of the A stars spectra were observed with ELODIE \citep{refId0} and SOPHIE \citep{2011SPIE.8151E..15P} spectrographs with a resolving power of 42\,000 and 75\,000, respectively. Details about the observations can be found in \cite{Gebran} and \cite{2019OAst...28...68K}. For the FGK stars, we have used 96 stellar spectra from the Spectroscopic Survey of Stars in the Solar Neighbourhood (S$^4$N, \citealt{2004A&A...420..183A}), analysed in \cite{S4n} having a resolving power of 50\,000. Like done in our previous studies \citep{S4n,Gebran}, variable or active stars, showing at time emission features of changing strength/amplitude were excluded.

For each resolving power, we used the corresponding trained NN model. To do that, the observations were corrected for the radial velocity shift using the classical cross correlation technique \citep{1979AJ.....84.1511T}. The spectra are then interpolated in the wavelength range used during the training, between 4450 and 5400 \AA\ for A stars and 5000 to 5400 \AA\ for FGK stars. The wavelength range of the S$^4$N observations is smaller than that of A stars. For that reason, a reduced database with smaller wavelength range was interpolated from the original one at 50\,000 resolving power, and was used for these specific FGK stars. Then all observed spectra were projected into their corresponding principal components and the first 50 data points were conserved for the prediction. 

\subsection{AFGK Stars}
\label{obs-A}
Predicted stellar parameters are depicted in Tabs.~\ref{results_A} and \ref{results_FGK} in the appendix. In these tables, the stellar parameters are represented with the median and closest values retrieved from Vizier catalogues using {\tt astroquery}\footnote{\url{https://astroquery.readthedocs.io}} \citep{vizier}\footnote{\url{https://arxiv.org/abs/1408.7026}}. 

\begin{figure}[!h]
    \centering
\includegraphics[scale=0.3]{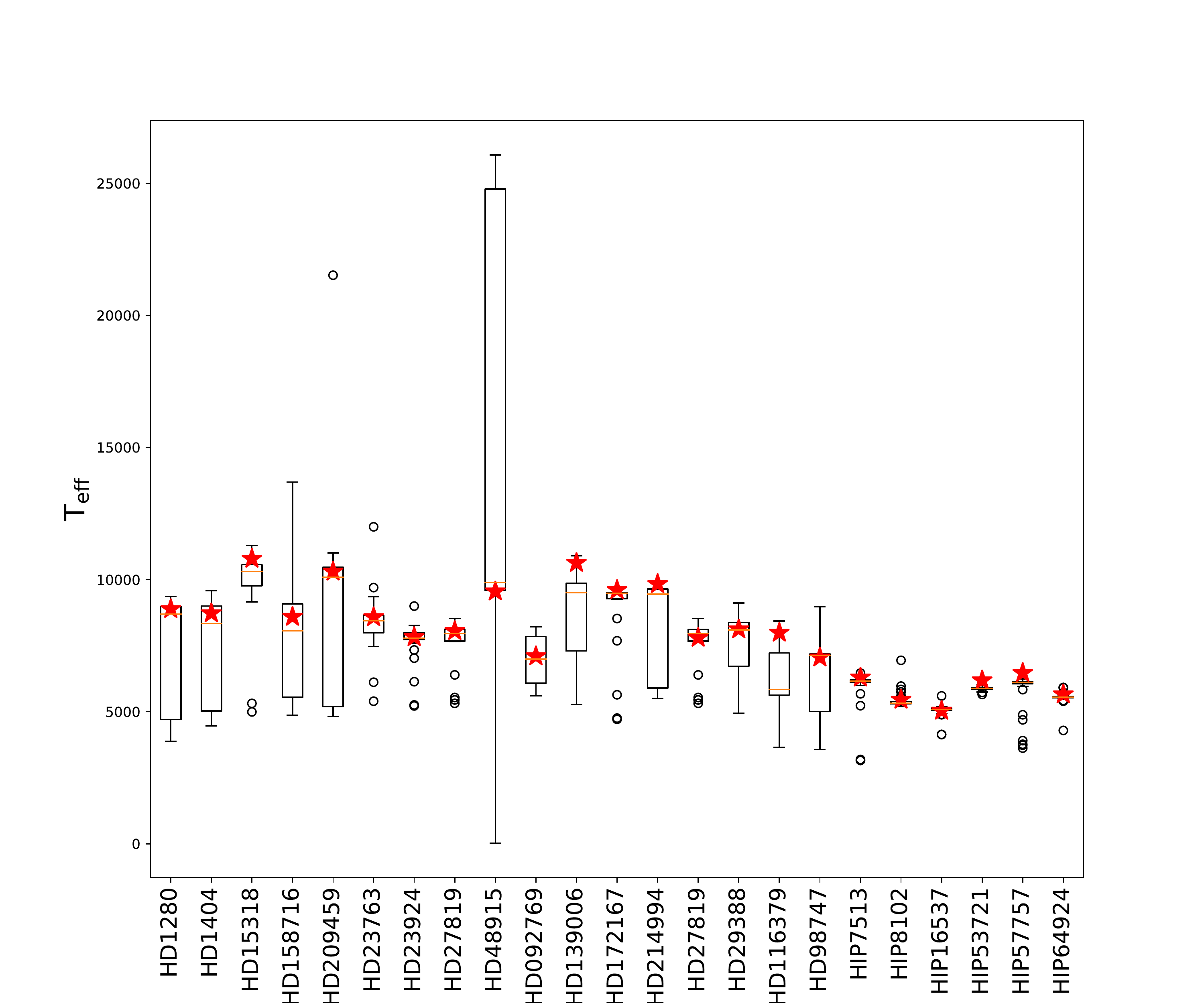}
    \caption{Comparison between our predicted effective temperatures (stars), and the values we obtained from available Vizier catalogues. The catalogued values are represented as classical boxplots. Objects we studied are listed along the horizontal axis.  The horizontal bar inside each box indicates the median (Q$_2$ value), while each box extends from first quartile, Q$_1$, to third quartile Q$_3$. Extreme values, still within a 1.5 times the interquartile range away
from either Q$_1$ or Q$_3$, are connected to the box with dashed lines. Outliers are denoted by a “o” symbol.}
    \label{boxplotTeff}
\end{figure}

\begin{figure}[!h]
    \centering
\includegraphics[scale=0.3]{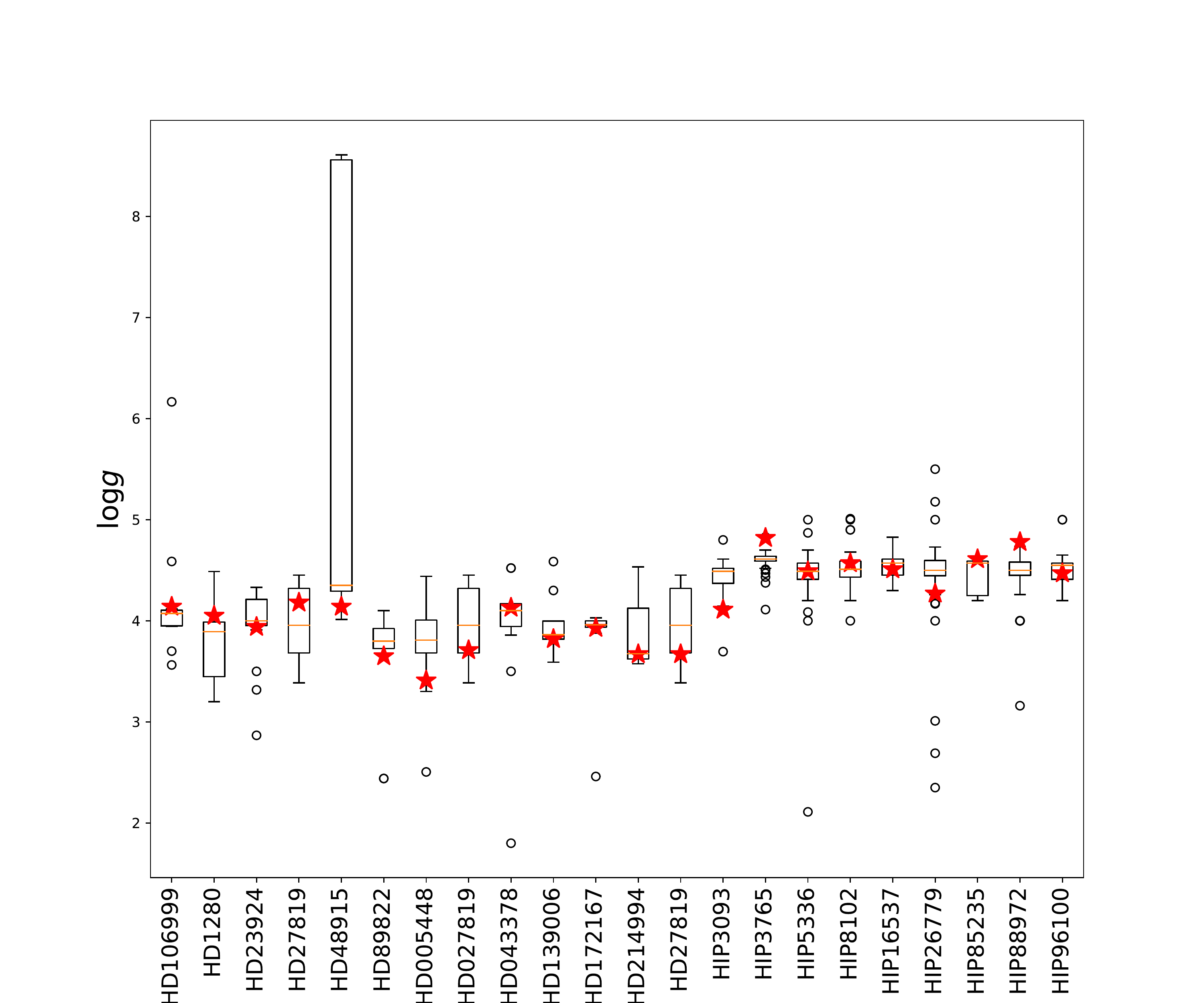}
    \caption{Same as Fig.~\ref{boxplotTeff} but for \logg.}
    \label{boxplotlogg}
\end{figure}

\begin{figure}[!h]
    \centering
\includegraphics[scale=0.3]{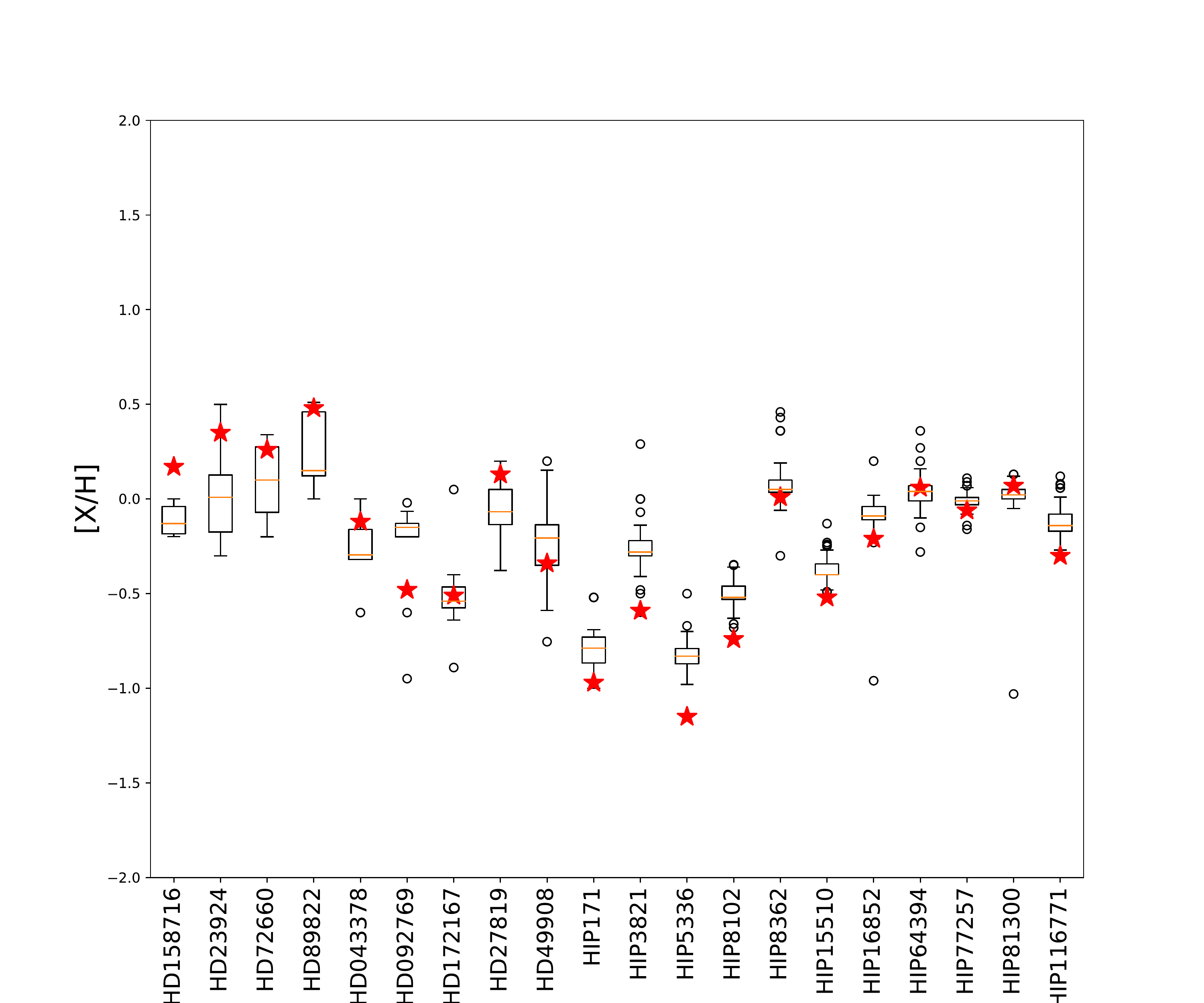}
    \caption{Same as Fig.~\ref{boxplotTeff} but for \met.}
    \label{boxplotmeta}
\end{figure}

\begin{figure}[!h]
    \centering
\includegraphics[scale=0.3]{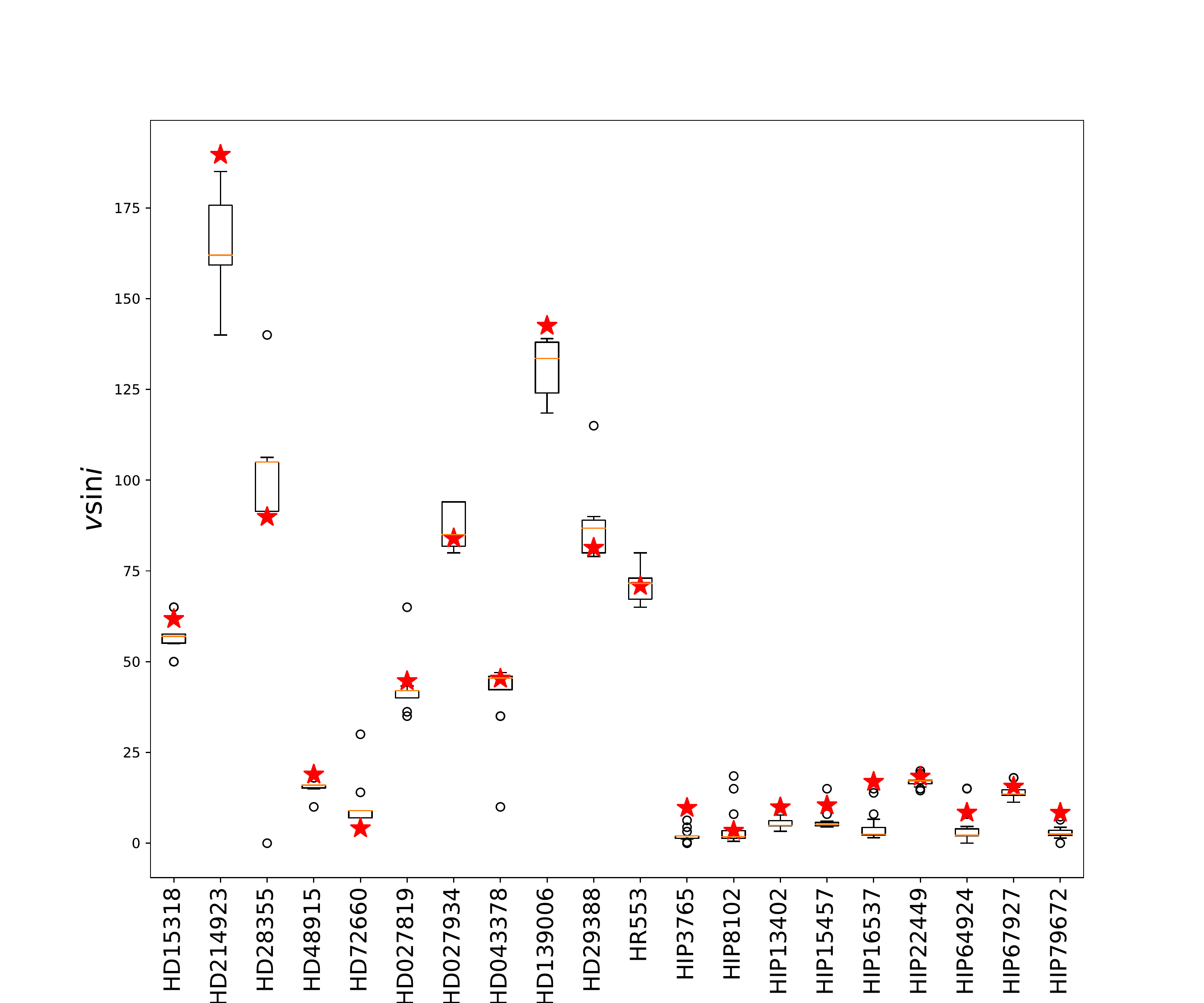}
    \caption{Same as Fig.~\ref{boxplotTeff} but for \vsini.}
    \label{boxplotvsini}
\end{figure}

Figure~\ref{boxplotTeff} shows the predicted effective temperature of a sample of stars as well as the range in the effective temperatures retrieved from the catalogues (box-plots) and the median. The selection of these stars was based on the number of values found in the literature. For A stars, we have selected the ones that have more than 20 different values in Vizier. As for FGK stars, we did the same with stars having more than 100 independent literature values for \Teff. Figures~\ref{boxplotlogg} to \ref{boxplotvsini} are similar to Fig.~\ref{boxplotTeff} but for \logg, \met, and \vsini. A stars in Fig.~\ref{boxplotlogg} have more than 10 catalogued values for \logg\ and F stars have more than 50.
For \met\ and \vsini, we have chosen the A stars that have more than 10 catalogued values for these two parameters. For FGK stars, we have selected the ones having more than 50 and 15 values for \met\ and \vsini, respectively.

A large spread exists in the literature values for all parameters. To estimate the accuracy of our results, we used a weighted mean approach similar to the one described in \cite{Gebran}. Quantitatively, and in order to give more weight to the catalogued values that have a large number of occurrence and a small spread in values, the dispersion and it corresponding standard deviation  for a stellar parameter $X$ are calculated as follows:
$$\Delta X = \frac{\sum_i w_i (X^{\mathrm{prediction}}-X^{\mathrm{median}})}{\sum_i w_i}$$
where
$$w_i = \frac{1}{\mathrm{IQR}_i}$$
$\mathrm{IQR}_i$ being the interquartile range defined as the difference between the third and first quartile of each set of values. The standard deviation is calculated as:

$$\sigma_X = \left[ \frac{\sum_i w_i (X^{\mathrm{prediction}}-X^{\mathrm{median}})^2}{\sum_i w_i}\right]^{\dfrac{1}{2}}$$ 

Stars having only one catalogued value for a specific parameter were not considered in the calculation of the dispersion. The results of the dispersion as well as the standard deviations are displayed in Tab.~\ref{tab:errors}. Catalogued values are all coming from different sources and each author uses a different technique (photometry, spectroscopy, spectrophotometry, asteroseismology$\ldots$). This leads to a large dispersion and a large deviation between our predicted values and the ones in the literature. A better way to estimate this dispersion is to do the comparison with the sample used in Figs.~\ref{boxplotTeff} to \ref{boxplotvsini}. This sample contains the stars having the largest number of independent catalogued values. The new dispersion and standard deviations are displayed in Tab.~\ref{tab:errors} depicted with the "lim" subscript. In that case, the dispersion reduces drastically, reaching an average of 150 K, 0.01 dex, 0.04 dex, and 3.0 km/s for \Teff, \logg, \met, and \vsini, respectively and with smaller standard deviation than in the case of the whole sample.

\begin{table}[]
    \centering
    \begin{tabular}{|c|c|c|c|c|}
    \hline
         &\Teff & \logg  & \met  & \vsini  \\
    &     (K) & (dex) & (dex) & (km/s) \\ \hline
     $\Delta$ & 160 & 0.40 & 0.15 & 12 \\
     $\sigma$ & 300 & 0.55 & 0.35 & 15 \\ \hline
     $\Delta_{\mathrm{lim}}$ & 150 & 0.01  & 0.04& 3.0 \\
     $\sigma_{\mathrm{lim}}$ & 250 & 0.15 & 0.14 &  5.5\\ \hline

    \end{tabular}
    \caption{Dispersion and standard deviation for the comparison between our predicted parameters and the catalogued ones. The last two rows deals with a limited catalogued sample, the one used to plot Figs.~\ref{boxplotTeff} to \ref{boxplotvsini}.}
    \label{tab:errors}
\end{table}

The dispersion found between our predicted \vsini\ for FGK and the ones in the literature are mainly explained by the fact that our \vsini\ include the macroturbulence effects whereas some of the authors derive both parameters separately (see for example \citealt{2004A&A...420..183A}). In the case of A stars, the \vsini\ values for very sharp-lined spectra (e.g., \vsini $\leq$ 5 km/s) should be considered as upper limits because macroturbulence effect is neglected. For moderate and fast rotators, macroturbulence has no significant effect on the line shape  \citep{2018PASJ...70...91T,2022arXiv220610986F}. When only considering A stars, $\Delta_{\mathrm{lim}}$ and $\sigma_{\mathrm{lim}}$ of Tab.~\ref{tab:errors} become 1.3 km/s and 5.0 km/s, respectively.  These results show that we are able to recover the stellar parameters of AFGK stars with good accuracy using our trained models. However, as explained in Paper I, the size of the database is very crucial for the convergence of the model as well as for the recovered accuracy of the stellar parameter. The size of the database depends on many factors, the spectral type of the stars, the wavelength range and the type of predicted parameters. We have used a database of $\sim$80\,000 spectra in our study but this number should be monitored.

\section{Discussion and Conclusion}
\label{discussion}
Two sources of errors should be assigned to the predicted stellar parameters. One relates to the model ($\sigma_{\mathrm{parameter}_m}$, Sec.~\ref{res}) and the other relates to how close the radiative transfer code ($i.e.$ \texttt{SYNSPEC48}, $\sigma_{\mathrm{parameter}_{\mathrm{rtc}}}$) represents the observations. $\sigma_{\mathrm{parameter}_{\mathrm{rtc}}}$ can be derived using a list of well studied stars with well established stellar parameters over a wide range in the HR Diagram. It is beyond the scope of this study, but we should be aware that this source of error could be wavelength dependant as each radiative transfer code uses a different line list with different atomic data. 

Model (i.e NN) and radiative transfer errors are independent and can be added in quadratic manner to find the total accuracy that we found in Sec.~\ref{obs-A}:
$$\sigma_{\mathrm{parameter}_{\mathrm{total}}}=(\sigma_{\mathrm{parameter}_m}^2 +\sigma_{\mathrm{parameter}_{\mathrm{rtc}}}^2 )^{\frac{1}{2}}$$

While comparing with the median values from literature, we found that \Teff\ is derived with an accuracy of 150 K, \logg\ with 0.01 dex, \met\ with 0.04 dex, and \vsini\ with 3.0 km/s. Some of these deviations are smaller than the errors found in the model (80 K for \Teff, 0.06 dex for \logg, 0.08 dex for \met, and 3.0 km/s for \vsini), but one should consider the accuracy of the model as a minimum limit for the stellar parameters and then calculate the total accuracy depending on the radiative transfer code specificity.

 We have used a large range of spectral type and found acceptable values for the accuracy. One could use a combination of stellar library with synthetic data adapted for each spectral type and luminosity range or a large database of observed stars with accurate stellar parameters. However, NN prove to be a fast (refer to Paper I for computational time) and accurate way to derive stellar parameters and can handle a large amount of data. 
 These results are very promising as they are less than the accuracy that are usually found with photometric techniques \citep{2005MSAIS...8..130S,JINMENG2021101568,2021ApJ...907...57G} or spectroscopic ones \citep{2018A&A...612A.111G,Gebran, 2019ApJ...879...69T,2019OAst...28...68K,2022A&A...657A..66T} or a combination of both \citep{2002A&A...392.1031A,2015A&A...582A..49H}.

In a future work, we will be testing the effect of specific spectral region on the stellar parameters. This will be done through autoencoders, a type of unsupervised learning technique, leading to a more "intelligent" and compact database construction. 

One straightforward application is the use of such a network in order to derive the stellar parameters of Gaia spectra \citep{2016A&A...595A...1G}. The radial velocity spectrometer (RVS, \citealt{2018A&A...616A...5C}) on board of Gaia will deliver medium resolution spectra (R$\sim$11\,000) in the CaII triplet region ($\lambda$ ranging from 8470 to 8710 \AA). Several millions of stars have their spectra available during the Data Release 3 (DR3, \citealt{2022arXiv220605870G,2022arXiv220610986F}).\\

\textbf{Acknowledgment}: We are very grateful to the referees of the paper for the useful remarks.\\

\textbf{Funding information}: This work was supported by the Neuhoff Summer Research Scholarship program at Saint Mary’s College.\\

\textbf{Author contributions}: All authors have accepted responsibility for the entire content of this manuscript and approve edits submission.\\

\textbf{Conflict of interest}: The authors state no conflict of inter-est.

\bibliographystyle{apalike}
\bibliography{biblio}

\appendix

\section{Tables}

\begin{table*}[!h]
\tiny
\begin{tabular}{|c|c|c|c|c|c|c|c|c|c|c|c|c|c|}
\hline
   HIP& HD& \Teff& $T_{\rm{eff}}^c$& $T_{\rm{eff}}^m$
   & \logg& $\logg g^c$& $\log g^m$
& \met & $[X/H]^c$& $[M/H]^m$    
& \vsini& $v\sin i^c$& $v \sin i^m$ \\
  & & (K)& (K)& (K)
   & (dex)& (dex)& (dex)
&  (dex) & (dex)& (dex)
& (km/s)& (km/s)& (km/s) \\
\hline
HIP100108 &HD193369   &10109&7718&10100&4.08&4.30&4.29&0.14&0.04&0.04&120.4&102.0&110.0\\
HIP102098 &HD197345   &7081&7823&7572&2.22&2.51&2.13&0.61&0.06&0.48&31.3&34.7&34.7\\
HIP102208 &HD199095   &10610&8934&10500&4.03&3.95&3.95&0.03&0.0&0.00&27.4&32.0&30.0\\
HIP103298 &HD199254   &7842&8145&7900&3.37&4.01&3.50&-0.24&-0.2&-0.40&165.8&148.0&159.0\\
HIP104139 &HD200761   &9959&9595&10001&3.74&4.11&4.00&0.28&0.26&0.26&145.6&104.0&130.0\\
HIP106297 &HD205117   &10091&9370&9800&4.02&3.90&4.00&-0.10&0.00&-0.10&138.5&83.5&90.0\\
HIP10670 &HD14055   &10174&9340&10772&4.17&4.08&4.19&-1.31&-0.58&-1.20&233.2&246.0&240.0\\
HIP10793 &HD14252   &8638&8380&8749&3.35&4.74&3.40&-0.01&-0.05&0.00&23.5&22.0&25.0\\
HIP108875 &HD209459   &10307&10093&10350&3.16&3.55&3.48&-0.95&-0.15&-0.42&2.0&11.0&3.8\\
HIP109521 &HD210715   &8099&7901&8200&4.01&4.13&4.09&0.15&-0.10&-0.01&155.5&138.0&144.0\\
HIP111123 &HD213320   &10826&10125&10864&3.79&4.05&3.76&1.11&0.41&0.49&22.1&21.0&23.0\\
HIP111169 &HD213558   &9852&9197&9840&4.23&4.00&4.20&0.19&-0.28&0.00&149.1&128.0&150.0\\
HIP112029 &HD214923   &10396&11032&10723&3.26&3.87&3.50&-0.24&0.00&-0.30&189.6&162.0&185.0\\
HIP112051 &HD214994   &9834&9452&9866&3.67&3.68&3.65&1.44&0.08&0.42&5.1&10.0&5.0\\
HIP114745 &HD219485   &10361&9396&10000&3.81&3.82&3.81&0.05&0.00&0.03&25.5&23.0&25.0\\
HIP11484 &HD15318   &10790&10308&10900&3.64&4.00&3.48&0.00&-0.10&-0.04&61.7&57.0&65.0\\
HIP12706 &HD016970   &8587&8407&8551&4.29&4.18&4.30&-0.13&-0.01&-0.01&192.0&186.0&190.0\\
HIP1366 &HD1280   &8887&8697&8857&4.05&3.89&4.00&0.39&-0.69&0.14&101.9&102.0&102.0\\
HIP1473 &HD1404   &8728&8332&8770&4.17&4.18&4.17&0.28&-0.09&0.05&138.3&119.0&123.0\\
HIP15154 &HD20149   &9661&8631&9800&3.43&3.65&3.50&0.05&0.00&0.06&22.1&23.0&23.0\\
HIP16322 &HD21686   &10199&9468&10000&3.61&4.00&3.67&-0.22&-0.2&-0.40&237.9&244.0&244.0\\
HIP17791 &HD23763   &8581&8441&8591&4.33&4.03&4.10&0.33&-0.14&0.01&139.5&104.0&110.0\\
HIP18717 &HD25175   &10460&8034&10500&3.44&3.83&3.59&-0.10&-0.51&-0.16&56.9&55.0&55.0\\
HIP19949 &HD26764   &10123&8215&9825&3.57&3.39&3.67&-0.66&-0.65&-0.65&241.7&229.0&249.0\\
HIP20542 &HD27819   &8056&7957&8050&4.18&3.96&4.11&0.44&-0.07&0.20&50.1&42.0&43.3\\
HIP20542 &HD27819   &7799&7957&7800&3.67&3.96&3.70&0.13&-0.07&0.17&44.5&42.0&43.3\\
HIP20635 &HD027934   &7737&8105&7800&3.35&3.81&3.40&0.35&-0.01&0.05&84.0&85.0&85.0\\
HIP20901 &HD28355   &7170&7705&7592&3.95&4.00&3.97&0.23&0.30&0.20&89.9&105.0&90.0\\
HIP20901 &HD028355   &6823&7705&6262&3.23&4.00&3.22&0.13&0.30&0.20&92.7&105.0&92.8\\
HIP21029 &HD28527   &7466&8086&7700&3.58&3.91&3.69&0.09&0.13&0.10&66.1&86.0&70.0\\
HIP21589 &HD29388   &8120&8100&8200&3.64&3.88&3.69&0.27&-0.05&0.13&81.3&86.8&80.0\\
HIP21683 &HD029488   &7731&7947&7800&3.76&3.80&3.80&0.23&0.09&0.10&137.7&128.0&128.3\\
HIP21683 &HD29488   &7687&7947&7614&3.46&3.80&3.67&0.02&0.09&0.09&141.9&128.0&128.3\\
HIP23497 &HD32301   &7795&7863&7800&3.66&3.88&3.80&0.53&-0.01&0.15&130.4&124.5&131.0\\
HIP24340 &HD33641   &7536&7560&7560&3.96&3.92&3.96&0.19&-0.03&0.12&94.8&84.5&92.0\\
HIP29997 &HD042818   &10830&9370&10834&4.02&4.16&4.03&-0.57&0.30&0.30&265.6&255.0&260.0\\
HIP30060 &HD043378   &10278&9120&9580&4.13&4.10&4.15&-0.12&-0.30&-0.10&45.3&45.5&45.0\\
HIP32104 &HD48097   &10091&7508&9463&4.31&4.10&4.34&0.00&-0.10&-0.01&120.0&101.0&110.0\\
HIP32349 &HD48915   &9554&9900&9580&4.14&4.35&4.20&0.45&0.33&0.50&18.9&16.0&18.0\\
HIP32921 &HD49908   &10035&5685&10200&3.48&3.52&3.52&-0.34&-0.21&-0.35&154.2&117.0&140.0\\
HIP36145 &HD58142   &9340&9462&9266&3.30&3.67&3.55&0.12&0.00&0.00&21.0&18.6&19.0\\
HIP41152 &HD070313   &8747&8038&8720&4.05&4.00&4.03&0.46&-0.48&-0.01&119.1&112.0&114.0\\
HIP42028 &HD72660   &9160&9513&9200&3.66&4.00&3.60&0.26&0.10&0.21&4.1&9.0&5.0\\
HIP4436 &HD5448   &8163&7118&8222&4.29&3.81&4.20&0.24&-0.17&0.10&68.7&75.0&69.3\\
HIP45493 &HD079439   &6751&7630&7450&4.09&4.04&4.10&-0.46&-0.04&-0.5&175.4&159.0&159.0\\
HIP50448 &HD88983   &7628&7890&7600&3.73&3.89&3.89&-0.19&-0.19&-0.18&126.1&114.0&133.0\\
HIP50933 &HD89822   &9661&10000&9741&3.65&3.80&3.66&0.48&0.15&0.46&3.9&10.0&4.6\\
HIP51200 &HD090470   &8241&7845&8337&4.01&4.20&4.20&0.06&-0.01&-0.01&125.2&90.0&110.0\\
HIP52422 &HD092769   &7100&6990&7600&4.42&4.13&4.30&-0.48&-0.15&-0.60&223.7&207.0&212.0\\
HIP5310 &HD006695   &8773&8304&8720&3.99&4.30&3.91&0.07&-0.20&-0.01&164.2&149.0&150.0\\
HIP53485 &HD94766   &7927&7908&7917&4.56&4.06&4.21&0.12&-0.05&0.00&94.7&85.0&85.0\\
HIP54326 &HD96399   &7414&6662&7400&3.62&3.72&3.40&-0.39&-0.49&-0.40&78.0&70.0&70.0\\
HIP54425 &HD96681   &7963&7638&7829&3.41&3.66&3.40&-0.03&-0.14&-0.01&79.1&80.0&80.0\\
HIP55263 &HD98377   &8813&8297&8800&4.68&4.01&4.13&-0.11&-0.10&-0.10&55.3&50.0&50.0\\
HIP5542 &HD6961   &7578&7962&7597&3.74&3.64&3.80&0.51&-0.20&0.11&103.3&103.0&103.0\\
HIP55488 &HD98747   &7056&7136&6992&4.03&3.91&4.15&-0.47&-0.12&-0.20&39.0&35.0&35.0\\
HIP56429 &HD100518   &7942&7637&7986&3.60&3.61&3.50&-0.13&-0.16&-0.02&8.2&11.2&8.0\\
HIP57743 &HD102841   &7173&7400&7181&4.41&3.70&4.55&-0.28&-0.30&-0.30&123.5&90.0&90.0\\
HIP59923 &HD106887   &7823&8291&7900&3.93&4.20&3.80&0.46&0.21&0.21&86.2&82.0&84.1\\
HIP59988 &HD106999   &8109&6519&8116&4.14&4.07&4.12&0.05&-0.21&0.08&50.4&47.7&51.4\\
HIP60327 &HD107655   &9153&8607&9281&3.78&4.00&3.97&0.79&-0.09&0.08&56.1&46.0&50.0\\
HIP62874 &HD112002   &8045&7716&8000&4.15&3.99&4.00&0.11&-0.45&0.10&54.7&50.0&50.0\\
HIP65304 &HD116379   &7993&5848&8000&3.82&4.25&3.80&0.08&-0.27&-0.02&89.2&80.0&80.0\\
HIP65466 &HD116706   &8907&8480&8909&3.92&3.93&3.93&0.35&-0.20&-0.01&56.3&54.0&55.0\\
HIP6686 &HD8538   &7945&7980&7980&3.72&3.61&3.73&0.01&-0.45&-0.11&127.6&110.0&123.0\\
HIP67004 &HD119537   &8740&8661&8661&3.97&3.99&3.99&0.20&-0.44&0.03&17.9&13.5&16.4\\
HIP73156 &HD132145   &9434&9230&9376&3.95&4.13&4.00&-0.24&0.00&0.00&15.3&15.0&15.0\\
HIP75043 &HD136729   &8295&8247&8279&3.88&4.19&3.85&-0.33&0.09&-0.30&161.3&159.0&161.0\\
HIP76267 &HD139006   &10635&9515&10900&3.82&3.86&3.82&-0.26&0.20&-0.01&142.5&133.5&139.0\\
HIP78554 &HD143894   &9246&8652&9226&3.97&3.93&3.93&0.28&0.38&0.38&149.2&128.0&130.0\\
HIP79332 &HD145647   &9674&7645&9560&3.93&3.41&3.95&-0.40&-0.30&-0.36&46.8&43.0&45.0\\
HIP84036 &HD155375   &8704&8477&8700&4.49&4.06&4.08&0.40&0.20&0.22&28.1&27.9&28.0\\
HIP84821 &HD157087   &8592&8185&8600&3.38&3.10&3.44&0.11&-0.05&0.00&8.9&15.0&12.0\\
HIP85666 &HD158716   &8593&8068&8600&3.82&4.26&3.82&0.17&-0.13&0.00&5.1&15.0&6.0\\
HIP8903 &--   &8107&8352&8061&3.88&3.94&3.90&0.34&0.08&0.16&70.8&71.6&71.6\\
HIP91262 &HD172167   &9608&9485&9657&3.93&3.96&3.93&-0.51&-0.54&-0.50&23.2&23.0&23.0\\
HIP92396 &HD174567   &10395&9208&10500&3.46&3.55&3.50&0.44&0.00&0.15&9.6&15.0&12.0\\
HIP93526 &HD176984   &9876&8723&9880&3.40&3.47&3.44&-0.10&-0.14&0.00&28.9&24.2&30.0\\
HIP9480 &HD012111   &7586&7700&7700&3.99&4.02&3.95&0.04&-0.31&-0.21&71.2&71.6&71.6\\
HIP97229 &HD186689   &7466&7906&7700&4.01&4.21&4.21&-0.10&-0.04&-0.05&32.5&31.0&31.0\\
HIP9977 &HD013041   &8420&8216&8309&3.72&3.86&3.77&-0.41&-0.45&-0.40&164.8&133.0&135.0\\
--   &HD23924   &7826&7782&7850&3.94&4.00&3.94&0.35&0.01&0.38&36.0&44.8&33.0\\

\hline
   \end{tabular}
    \caption{Predicted values for \Teff, \logg, \met, and \vsini\ for A stars with the median and closest values from Vizier catalogue.}
    \label{results_A}
\end{table*}

\begin{table*}[!h]
\tiny
\begin{tabular}{|c|c|c|c|c|c|c|c|c|c|c|c|c|}
\hline
   Star& \Teff& $T_{\rm{eff}}^c$& $T_{\rm{eff}}^m$
   & \logg& $\logg g^c$& $\log g^m$
& \met & $[X/H]^c$& $[M/H]^m$    
& \vsini& $v\sin i^c$& $v \sin i^m$ \\
   ID& (K)& (K)& (K)
   & (dex)& (dex)& (dex)
&  (dex) & (dex)& (dex)
& (km/s)& (km/s)& (km/s) \\
\hline
HIP10138 &5194&5188&5195&4.91&4.56&4.91&0.01&-0.23&-0.09&7.0&2.3&3.9 \\
HIP102422 &4937&4971&4940&2.54&3.40&2.99&0.30&-0.18&0.13&4.7&3.4&4.8 \\
HIP105858 &5909&6159&5910&3.47&4.35&3.92&-1.10&-0.67&-0.84&24.0&3.7&10.0 \\
HIP10644 &5871&5702&5845&3.92&4.29&3.92&-0.75&-0.43&-0.58&31.6&4.7&10.0 \\
HIP10798 &5186&5373&5286&4.52&4.61&4.53&-0.84&-0.47&-0.80&8.8&2.7&3.6 \\
HIP109176 &6664&6479&6693&3.72&4.23&4.02&-0.21&-0.10&-0.19&20.5&6.2&10.0 \\
HIP110109 &6106&5850&6019&3.74&4.39&4.13&-0.70&-0.21&-0.44&8.3&2.0&2.7 \\
HIP114622 &4755&4829&4749&2.12&4.50&2.59&0.92&0.05&0.20&9.0&2.0&8.0 \\
HIP116771 &6273&6186&6279&3.58&4.12&3.75&-0.30&-0.14&-0.27&11.1&6.7&10.0 \\
HIP12777 &6328&6264&6329&3.66&4.32&3.22&-0.16&-0.01&-0.15&12.6&8.9&10.2 \\
HIP12843 &5523&6371&6144&3.74&4.29&4.00&0.15&0.05&0.15&25.0&25.6&25.0 \\
HIP13402 &5171&5180&5170&4.54&4.56&4.55&0.62&0.08&0.21&9.9&4.9&10.0 \\
HIP14632 &6340&5963&6045&3.61&4.16&3.35&0.66&0.09&0.29&8.5&4.3&10.0 \\
HIP14879 &6160&6170&6165&3.50&3.95&3.57&-0.40&-0.21&-0.35&10.5&4.4&7.3 \\
HIP15330 &6071&5720&5854&4.28&4.53&4.30&-0.84&-0.22&-0.34&12.2&2.7&3.0 \\
HIP15371 &6155&5866&6066&3.96&4.48&4.22&-0.77&-0.23&-0.34&9.1&2.6&3.0 \\
HIP15457 &5908&5718&5908&3.60&4.50&4.33&-0.16&0.06&-0.16&10.4&5.2&8.0 \\
HIP15510 &6198&5401&6041&4.51&4.45&4.50&-0.52&-0.40&-0.49&7.0&1.5&4.0 \\
HIP1599 &6234&5957&6151&3.70&4.42&4.02&-0.43&-0.19&-0.45&16.4&3.0&15.0 \\
HIP16537 &5039&5084&5034&4.51&4.57&4.51&0.27&-0.11&0.06&16.9&2.5&15.0 \\
HIP16852 &6183&5997&6200&3.47&4.09&3.85&-0.21&-0.09&-0.21&8.3&4.3&8.0 \\
HIP171 &5853&5438&5798&4.40&4.38&4.40&-0.97&-0.79&-0.98&15.1&3.0&5.0 \\
HIP17378 &4734&5037&4750&2.35&3.77&3.27&0.36&0.10&0.25&24.7&2.3&15.0 \\
HIP17420 &4957&4979&4957&3.86&4.57&4.41&0.36&-0.11&0.10&9.4&3.0&5.7 \\
HIP2021 &6042&5848&5924&3.42&3.95&3.45&0.00&-0.09&0.00&9.2&3.3&5.0 \\
HIP22263 &6300&5834&6131&3.72&4.49&4.30&-0.48&0.01&-0.19&13.0&3.2&6.4 \\
HIP22449 &5857&6424&5820&3.52&4.29&3.77&0.02&0.00&0.02&18.3&17.2&18.5 \\
HIP23311 &4790&4790&4790&2.09&4.55&4.23&1.18&0.28&0.44&21.4&2.0&5.2 \\
HIP23693 &5838&6153&5727&3.66&4.44&4.06&-0.32&-0.17&-0.34&17.7&15.4&17.3 \\
HIP24813 &6167&5858&5979&3.67&4.20&3.98&0.51&0.05&0.26&3.7&2.0&3.1 \\
HIP26779 &5301&5243&5300&4.27&4.50&4.26&0.49&0.09&0.21&15.5&2.5&5.4 \\
HIP27072 &6381&6306&6384&3.64&4.31&3.99&-0.23&-0.05&-0.22&12.7&7.7&10.4 \\
HIP27913 &5892&5949&5895&3.74&4.44&4.21&-0.40&-0.03&-0.18&13.0&8.9&10.7 \\
HIP29271 &5628&5569&5621&3.70&4.43&4.20&0.77&0.10&0.25&18.4&1.8&2.3 \\
HIP3093 &5018&5221&5024&4.11&4.49&4.15&0.51&0.15&0.26&7.6&1.2&8.0 \\
HIP37279 &6770&6596&6775&3.47&4.00&3.74&-0.27&-0.01&-0.29&10.7&5.5&10.1 \\
HIP37349 &4812&4932&4826&2.74&4.60&2.68&0.83&-0.01&0.09&6.6&3.8&5.6 \\
HIP3765 &5024&4978&5020&4.82&4.61&4.82&0.06&-0.24&-0.04&9.7&2.0&6.3 \\
HIP3821 &6022&5925&6034&3.65&4.40&4.00&-0.59&-0.28&-0.60&10.5&2.8&9.2 \\
HIP40693 &5428&5402&5428&3.79&4.48&3.66&0.16&-0.03&0.14&9.0&2.0&6.7 \\
HIP4148 &4688&4952&4822&3.50&4.61&4.49&0.32&-0.11&0.00&8.7&1.8&4.5 \\
HIP41926 &5080&5243&5155&4.69&4.56&4.68&-0.73&-0.40&-0.48&8.8&2.7&6.8 \\
HIP42438 &5765&5876&5759&3.66&4.47&4.40&-0.35&-0.06&-0.29&13.8&10.0&13.2 \\
HIP42808 &5018&4969&5005&4.73&4.60&4.66&0.75&-0.03&0.10&8.9&3.8&9.6 \\
HIP46853 &6217&6336&6225&3.29&3.87&3.50&-0.42&-0.16&-0.31&12.3&8.6&10.0 \\
HIP51459 &6301&6156&6301&3.57&4.39&3.96&-0.32&-0.13&-0.28&9.3&4.3&10.0 \\
HIP5336 &5335&5316&5336&4.49&4.49&4.49&-1.15&-0.83&-0.98&19.3&5.4&15.0 \\
HIP53721 &6186&5882&6140&3.70&4.30&4.07&0.52&0.01&0.31&6.2&2.8&5.6 \\
HIP544 &5546&5481&5551&3.44&4.55&3.99&0.64&0.12&0.22&8.9&4.1&6.2 \\
HIP56452 &5128&5158&5129&4.80&4.56&4.68&-0.92&-0.38&-0.57&8.1&3.5&6.7 \\
HIP56997 &5605&5507&5609&3.71&4.54&3.45&-0.48&-0.05&-0.50&33.5&2.4&15.0 \\
HIP57443 &5982&5629&5970&4.21&4.44&4.21&-0.83&-0.29&-0.66&10.3&0.7&3.0 \\
HIP57757 &6470&6109&6246&3.55&4.10&3.86&0.52&0.13&0.33&8.3&4.0&10.0 \\
HIP58576 &5332&5510&5361&3.37&4.40&3.65&0.94&0.25&0.35&8.4&2.0&5.2 \\
HIP61317 &6063&5881&6061&3.70&4.39&3.38&-0.56&-0.19&-0.39&2.1&2.8&2.1 \\
HIP61941 &5502&6875&5433&3.64&4.26&3.88&-0.40&-0.09&-0.44&28.6&28.3&29.7 \\
HIP64241 &5687&6343&5250&3.64&4.09&3.99&0.01&-0.23&0.0&20.9&19.9&20.5 \\
HIP64394 &6517&6009&6225&3.72&4.40&4.24&0.06&0.04&0.06&10.7&4.4&10.0 \\
HIP64924 &5660&5558&5660&3.75&4.40&3.50&-0.17&-0.01&-0.13&8.4&2.2&8.0 \\
HIP67927 &6076&6047&6078&3.49&3.78&3.53&0.82&0.25&0.47&15.5&13.5&15.4 \\
HIP68184 &4776&4831&4792&2.20&4.55&4.38&1.00&0.12&0.33&25.4&1.3&9.0 \\
HIP71681 &5203&5551&5203&4.11&4.31&4.16&0.57&0.21&0.27&11.3&2.7&4.5 \\
HIP72659 &5616&5483&5595&4.23&4.56&4.37&-0.67&-0.14&-0.83&15.7&4.6&16.0 \\
HIP72848 &5290&5260&5291&4.10&4.53&4.11&0.60&0.08&0.14&9.5&4.5&6.3 \\
HIP73695 &6160&5495&6200&3.78&4.23&4.10&-0.54&-0.30&-0.42&6.2&3.7&3.7 \\
HIP7513 &6296&6155&6269&3.65&4.13&3.90&0.24&0.09&0.19&13.7&9.6&11.9 \\
HIP77257 &6206&5901&6131&3.57&4.15&4.00&-0.06&-0.01&-0.05&8.3&3.1&10.0 \\
HIP7751 &4969&5043&4970&4.58&4.63&4.61&-0.27&-0.20&-0.26&8.5&3.9&6.8 \\
HIP77952 &5492&7107&5377&2.77&4.16&3.76&0.11&-0.25&-0.20&79.2&75.0&75.0 \\
HIP78072 &6117&6278&6146&3.67&4.13&3.91&-0.36&-0.18&-0.32&13.9&10.0&11.9 \\
HIP78775 &5321&5294&5321&4.72&4.58&4.71&-0.85&-0.67&-0.76&6.6&2.0&7.0 \\
HIP7918 &6232&5880&6179&3.73&4.30&4.10&0.68&0.0&0.2&7.1&3.2&5.0 \\
HIP79190 &5033&5060&5024&4.91&4.55&4.66&-0.18&-0.37&-0.21&8.7&1.6&5.0 \\
HIP79672 &6213&5799&6053&3.71&4.43&4.16&-0.32&0.04&-0.29&8.3&2.5&8.3 \\
HIP7981 &5154&5201&5155&3.34&4.50&3.25&0.13&-0.04&0.12&10.4&1.7&10.0 \\
HIP80337 &6525&5882&6060&3.68&4.50&4.40&-0.30&0.03&-0.19&17.5&1.6&3.9 \\
HIP80686 &6417&6090&6459&3.68&4.45&4.24&-0.30&-0.08&-0.19&10.8&3.2&3.3 \\
HIP8102 &5456&5330&5459&4.57&4.51&4.57&-0.74&-0.52&-0.68&3.4&1.8&3.5 \\
HIP81300 &5087&5272&5080&4.15&4.57&4.39&0.07&0.02&0.07&8.0&2.0&4.1 \\
HIP81693 &5994&5764&5906&3.20&3.74&3.53&0.77&0.02&0.10&8.4&4.3&10.0 \\
HIP8362 &5257&5374&5257&3.56&4.54&4.30&0.01&0.05&0.01&7.2&1.3&10.0 \\
HIP84405 &5009&5089&5007&4.80&4.60&4.64&-0.48&-0.23&-0.38&7.7&2.5&5.1 \\
HIP84720 &5285&5209&5273&4.76&4.53&4.61&-0.77&-0.34&-0.46&8.6&1.9&4.5 \\
HIP84862 &6270&5703&6079&3.90&4.26&3.80&-0.70&-0.37&-0.79&2.9&1.7&3.0 \\
HIP85235 &5072&5290&5194&4.61&4.57&4.61&-0.74&-0.44&-0.52&5.0&1.3&3.4 \\
HIP86036 &6490&5893&6077&3.65&4.39&4.13&0.11&-0.03&0.08&13.5&4.5&6.0 \\
HIP86400 &4863&4883&4864&2.24&4.52&4.30&0.50&-0.08&0.17&13.7&2.5&4.1 \\
HIP86974 &5361&5508&5342&2.86&3.97&3.72&0.82&0.23&1.29&7.8&3.9&8.0 \\
HIP88601 &5174&5250&5182&3.82&4.54&4.30&0.30&-0.01&0.19&11.0&3.5&13.0 \\
HIP88972 &5034&5000&5035&4.78&4.50&4.74&0.37&-0.17&0.07&5.0&2.1&4.1 \\
HIP91438 &5918&5636&5884&4.26&4.49&4.25&-0.90&-0.24&-0.35&34.1&2.8&4.0 \\
HIP96100 &5261&5271&5260&4.47&4.55&4.49&-0.70&-0.22&-0.43&7.2&2.3&6.7 \\
HIP97944 &5052&4767&5081&2.13&4.20&2.00&0.93&-0.03&0.38&33.7&2.0&10.2 \\
HIP98036 &5082&5100&5082&2.60&3.55&3.04&0.13&-0.15&-0.04&6.6&2.5&4.6 \\
HIP99461 &4949&4971&4952&4.80&4.55&4.73&-0.10&-0.51&-0.21&9.9&1.8&3.9 \\
HIP99825 &5233&5091&5179&4.14&4.51&4.37&0.46&0.00&0.14&18.0&2.0&4.3 \\
\hline
   \end{tabular}
    \caption{Predicted values for \Teff, \logg, \met, and \vsini\ for FGK stars with the median and closest values from Vizier catalogue.}
    \label{results_FGK}
\end{table*}

\end{document}